\documentclass[letter,superscriptaddress,twocolumn, pra,showkeys,showpacs]{revtex4}

	\usepackage{amsmath}
	\usepackage{makeidx}
	\usepackage{amsfonts}
	\usepackage[ansinew]{inputenc}
	\usepackage[usenames,dvipsnames]{pstricks}
    \usepackage{graphicx}
    \usepackage{float}
    \usepackage{subfigure}
	\usepackage{pst-grad} 
	\usepackage{pst-plot} 
	\usepackage{sidecap}
	\usepackage[makeroom]{cancel}
	\usepackage{textcomp}

	\usepackage[colorlinks,hyperindex]{hyperref}
	
	\hypersetup
	{
		colorlinks,%
		citecolor=black,%
		linkcolor=black,%
		urlcolor=black,%
	}

	\newcommand{\ket}[1]{\left| #1 \right\rangle}
	\newcommand{\bra}[1]{\left\langle #1 \right|}

\makeindex

\begin{document}

\title{Quantum Zeno and Anti-Zeno Effects on the Entanglement Dynamics of Qubits Dissipating into a Common and non-Markovian Environment}

\author{A. Nourmandipour}
\email{anoormandip@stu.yazd.ac.ir}
\affiliation{Atomic and Molecular Group, Faculty of Physics, Yazd University,  89195-741 Yazd, Iran}
\author{M. K. Tavassoly}
\email{mktavassoly@yazd.ac.ir}
\affiliation{Atomic and Molecular Group, Faculty of Physics, Yazd University,  89195-741 Yazd, Iran}
\affiliation{Photonic Research Group, Engineering Research Center, Yazd University,  89195-741 Yazd, Iran}
\author{M. A. Bolorizadeh}
\email{ma.bolorizadeh@kgut.ac.ir}
\affiliation{Physics and Photonics Department, Kerman Graduate University of Technology, Kerman, Mahan, Iran}
\affiliation{International Center for Science, High Technology and the Environmental Sciences, Mahan, Iran}

\date{today}

\begin{abstract}
We investigate the quantum Zeno and anti-Zeno effects on pairwise entanglement dynamics of a collective of non-interacting qubits which have been initially prepared in a Werner state and are off-resonantly coupled to a common and non-Markovian environment. We obtain the analytical expression of the concurrence in the absence and presence of the non-selective measurements. In particular, we express our results in the strong and weak coupling regimes and examine the role of the system size, and the effect of the detuning from the cavity field frequency on the temporal behaviour of the pairwise entanglement. We show that, the detuning parameter has a positive role in the protection of entanglement in the absence of the measurement for weak coupling regime. We find that for the values of detuning parameter less than the cavity damping rate, the quantum Zeno effect is always dominant, while for the values greater than the cavity damping rate, both Zeno and anti-Zeno effects can occur, depending on the measurement intervals.  We also find that the anti-Zeno effect can occur in the pairwise entanglement dynamics in the absence and presence of the detuning in the strong  coupling regime. \\
\end{abstract}


\maketitle

\section{Introduction}
One of the remarkable features of quantum theory which has no classical counterpart, is the idea of entanglement \cite{Horodecki2009QuantumEntanglement}. This strange phenomenon is an important resource for quantum information applications \cite{Ekert1991Cryptography,Braunstein1995,Mattle1996,Abdi2012,Muarao1999}.
Although the unavoidably interaction between  real quantum systems with their surrounding environment may cause the loss of entanglement stored in those systems, it has been shown that, under special conditions the common environment can have a constructive role in establishing and preserving the entanglement between subsystems even without any interaction among them \cite{Plenio2002,Nourmandipour2015,Memarzadeh2013,Maniscalco2008,Zhang2012,Nourmandipour2016}. Recently, the problem of two qubits in a common bath without using the rotating-wave, Born, and Markovian approximations has been investigated in \cite{Ma2012}.  Altogether, it seems quite logical to find a way to protect entanglement under contamination of the environment. Accordingly, many attempts have been devoted to fight against the deterioration of entanglement  under the impact of environment \cite{Kim2012,Ghanbari2014,Wang2013}.

In this regard, the quantum Zeno effect (QZE) is a promising implementation to protect entanglement from decoherence induced by the environment \cite{Misra1977,Koshino2005}. It relies upon the inhibition of the evolution of an unstable quantum system by frequent measurements during a defined period of time. This can be done when the state of the system evolves only in a multidimensional subspace, namely the Zeno subspace \cite{Facchi2002}. The two-particle quantum Zeno dynamics with a type of nondeterministic collective measurement with  specific outcomes has been investigated in \cite{Wang2008}. Cao et.al., have investigated the quantum Zeno effect for a qubit inside either a low- or high-frequency bath beyond the rotating wave approximation \cite{Cao2010}. It should be noticed that, the quantum Zeno effect for the entanglement is not straightforwardly predictable and in some cases, the repeated measurements can accelerate the decay of the encoded entanglement. This is quantum anti-Zeno effect (QAZE) \cite{Kofman2000,Facchi2001,Zheng2008,Ai2013}. Beside such theoretical studies, these two phenomena have been experimentally observed in many works \cite{Fischer2001,Vijay2012,Schindler2013}.

Here, we investigate the pairwise entanglement dynamics of an arbitrary number of qubits off-resonantly coupled to a common and non-Markovian environment for both weak and strong couplings corresponding to the bad and good cavity limits, respectively. We obtain the exact dynamics of  pairwise entanglement as a function of the environment correlation time when the qubits are initially in a Werner state for both coupling regimes. We then provide a series of nonselective measurements to check whether the system is still in its initial state after each measurement, and obtain the relevant pairwise concurrence after these $N$ measurements.

We show that, in the absence of nonselective measurements, on average, the detuning parameter has a positive role in surviving of entanglement in both coupling regimes. However, in the good cavity limits and for small detuning, this parameter can also have a destructing role in some regions of time. In the absence of detuning and in the bad cavity limit, the quantum Zeno effect is dominant for any value of system size $n$ and measurement time $T$. But, in the good cavity limit, both Zeno and anti-Zeno effects can be occurred depending on the system size and the measurement time intervals.

The rest of paper is organized as follows: In Sec. \ref{sec.model} we introduce the relevant Hamiltonian and by considering the initial state of qubits as a Werner state, we obtain the explicit form of the wave function of the system at any time $t$. In Sec. \ref{sec:OED} we obtain the expression for the concurrence between two arbitrary qubits in the absence and  presence of the nonselective measurements. Section \ref{sec:RES} deals with the investigation of pairwise entanglement in various situations. Finally, the paper ends with a summary and conclusion in Sec. \ref{sec:CON}.

\section{Model and Time Evolution of the System}\label{sec.model}

We consider a system consists of $n$ non-interacting qubits with associated Hilbert space ${\cal H}\simeq{\cal C}^{2\otimes n}$ dissipating into a common environment which in general can be considered outside of the Markovian limit. We assume  $\left\lbrace \ket{0},\ket{1}\right\rbrace ^{\otimes n}$ be the orthonormal basis in which, $\ket{0}$ ($\ket{1}$) is the ground (excited) single qubit state.

In \cite{Nourmandipour2015,nourmandipour2015novel} by assuming that the leakage of photons into a continuum state is the source of dissipation, we have shown that how a dissipative cavity can be modelled as a high-Q cavity in which the qubits interact with the cavity field and the cavity field itself interacts with an external field which can be considered as a set of continuum harmonic oscillators. More specifically, the corresponding Hamiltonian in the dipole and rotating-wave approximations and in units of $\hbar=1$ can be written as  \cite{Nourmandipour2015} 
\begin{equation}
\label{eq:Horigin}
\begin{aligned}
\hat{H}&=\dfrac{1}{2}\sum_{i=1}^{n}\omega_{\text{qb}_i}\hat{\sigma}_{z}^{(i)} +\omega_{0}\hat{a}^{\dagger}\hat{a} + \int_0^{\infty}\! \eta \hat{B}^{\dagger}(\eta)\hat{B}(\eta) \, \mathrm{d}\eta  \\
&+ \int_0^{\infty}\! \left(   G(\eta)\hat{a}^{\dagger}\hat{B}(\eta)\ + \ \text{H.c.} \right)  \, \mathrm{d}\eta  \\
&+ \sum_{i=1}^{n}\left( g_i\hat{\sigma}_+^{(i)}\hat{a}+ \text{H.c.} \right), 
\end{aligned}
\end{equation} 
where,  $\hat{\sigma}_{+}^{(i)}$ ($\hat{\sigma}_{-}^{(i)}$) and $\hat{\sigma}_{z}^{(i)}$ are the raising (lowering) and inversion population operators of the $i$th qubit with corresponding resonance frequency $\omega_{\text{qb}_i}$, $g_{i}$ is the coupling constant between $i$th atom and the cavity field and $\hat{a}$ ($\hat{a}^{\dagger}$) and $\omega_0$ are the annihilation (creation) operator and frequency of the cavity field, respectively. $G(\eta)$ is the coupling coefficient which in general, is a function of frequency that connects the external world to the cavity, and $\hat{B}^{\dagger}(\eta)$ and $\hat{B}(\eta)$ are the creation and annihilation operators of the surrounding environment in the mode $\eta$ obeying the commutation relation $ \left[ \hat{B}(\eta),\hat{B}^{\dagger}(\eta^{'}) \right]  =\delta(\eta-\eta^{'})$. In performing the Hamiltonian (\ref{eq:Horigin}), we assumed that there is no interaction among the qubits, and their positions have not been considered in our model. Therefore, it is possible to reduce the border effects.

\begin{figure}[ht]
\includegraphics[width=0.4\textwidth]{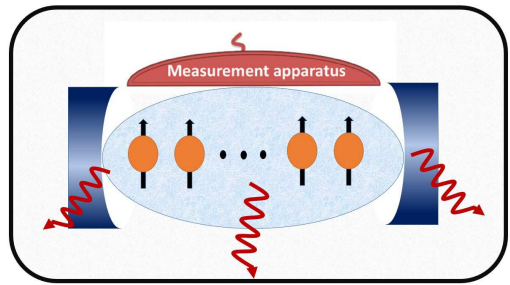}
\caption{Pictorial representation of the system under study.} \label{Fig1}
\end{figure}

The surrounding medium can be assumed to have  a narrow bandwith through which only a particular mode of the cavity may be excited. This assumption allows one to take the coupling coefficient $G(\eta)$ as a constant and equal to $\sqrt{\kappa/\pi}$ and also to diagonalize the Hamiltonian using dressed operators  \cite{Fano1961} $\hat{A}(\omega)=\alpha(\omega)\hat{a}+\int\! \beta(\omega,\eta)\hat{B}(\eta) \, \mathrm{d}\eta$ \cite{Nourmandipour2015}, in which
\begin{subequations}
\begin{eqnarray}
\alpha(\omega)&=&\frac{\sqrt{\kappa/\pi}}{\omega-\omega_0+i\kappa}, \label{eq:alpha} \\
\beta(\omega,\eta)&=& \sqrt{\kappa/\pi}\alpha(\omega)\left[  P\dfrac{1}{\omega-\eta}+\dfrac{\omega-\omega_0}{\kappa/\pi}\delta(\omega-\eta)\right], \label{eq:beta}
\end{eqnarray}
\end{subequations}
where $P$ refers to the principal value. The parameter $\kappa$ is the decay rate factor of the cavity \cite{Nourmandipour2015}. There is no need to check that the new operators $\hat{A}(\omega)$ and $\hat{A}^{\dagger}(\omega)$ satisfy the commutation relation $
 \left[ \hat{A}(\omega),\hat{A}^{\dagger}(\omega^{'}) \right]  =\delta(\omega-\omega^{'}) $.
Using this approach, one can treat the system outside of the Markovian regime. 
Therefore, the Hamiltonian (\ref{eq:Horigin}) in terms of the dressed operators  $\hat{A}(\omega)$ becomes 
\begin{equation}
\label{H2}
\begin{aligned}
\hat{H}&=\dfrac{\omega_{\text{qb}}}{2}\sum_{i=1}^{n}\hat{\sigma}_{z}^{(i)} + \int\! \omega \hat{A}^{\dagger}(\omega)\hat{A}(\omega) \, \mathrm{d}\omega  \\
&+ g\sum_{i=1}^{n}\int\!\left(\hat{\sigma}_+^{(i)}\alpha^{*}(\omega)\hat{A}(\omega)+\text{H.c.}\right) \, \mathrm{d}\omega,
\end{aligned}
\end{equation}
in which, we assumed that the resonant frequency of all qubits be same (namely, $\omega_{\text{qb}}$) and also the coupling constant between qubits and the cavity field be real and equal for all qubits (namely, $g$).  The obtained Hamiltonian clearly implies that the qubits are dissipating in a common environment (see Fig. \ref{Fig1}). 

In the continuation, we restrict ourselves to the case in which only one excitation of the whole system is considered. Therefore, the initial state of the form
\begin{equation}
\ket{\psi_0}=\ket{W}\ket{\boldsymbol{0}}_{R}
\label{eq:initialwerner}
\end{equation}
evolves after a time $t$ into a state
\begin{equation}
\ket{\psi(t)}={\cal E}(t)e^{i\omega_{\text{qb}}t}\ket{W}\ket{\boldsymbol{0}}_{R}
+\int\! \Lambda_{\omega}(t)e^{i\omega t}\ket{G}\ket{1_{\omega}} \, \mathrm{d}\omega,
\label{eq:statewerner}
\end{equation}
in which
\begin{equation}
\ket{W}:=\dfrac{1}{\sqrt{n}}\sum_{k=0}^{n}\ket{1_k}
\label{eq:wernerstate}
\end{equation}
is the Werner state where $\ket{1_k}\equiv\ket{0_1,\cdots,1_k,\cdots,0_n}$ implies that only $k$th qubit is in the excited state while all the others are in ground state and $\ket{\boldsymbol{0}}_{R}=\hat{A}(\omega)\ket{1_{\omega}}$ is the multi-mode vacuum state, where $\ket{1_{\omega}}=\hat{A}^{\dagger}(\omega)\ket{\boldsymbol{0}}_{R}$ is the multi-mode state representing one photon at frequency $\omega$ and vacuum state in all other modes and $\ket{G}:=\ket{0}^{\otimes n}$. In relation (\ref{eq:statewerner})
\begin{equation}
\left| {\cal E}(t)\right| ^2\equiv\text{P}_0(t)=\left|  \left\langle\psi_0 | \psi(t)\right\rangle \right|^2
\label{eq:suramp}
\end{equation}
is the survival probability of the initial state. By inserting $\ket{\psi(t)}$ into the time-dependent Schrödinger equation and after a long but straightforward manipulations, one may get the following closed equation for ${\cal E}(t)$
\begin{equation}
\label{eq:diffu}
\dot{{\cal E}}(t)=-\int_{0}^{t}\! f(t-t_1){\cal E}(t_1) \, \mathrm{d}t_1,
\end{equation}
in which the correlation function $f(t-t_1)$ is related to the spectral density $J(\omega)$ of the environment as
\begin{equation}
f(t-t_1)=\int\! \, \mathrm{d}\omega J(\omega) e^{-i\delta_{\omega}(t-t_1)}  , \label{eq:f}
\end{equation}
where $\delta_{\omega}=\omega-\omega_{\text{qb}}$ and according to Eq. (\ref{eq:alpha}) the Lorentzian spectral density reads as
\begin{equation}
\label{eq:spcden}
J(\omega)\equiv ng^2|\alpha(\omega)|^2=\dfrac{1}{\pi}\dfrac{ng^2\kappa}{ (\omega-\omega_0)^2+\kappa^2}.
\end{equation}
The parameter $\kappa$  is the spectral width and therefore describes the cavity losses (photon escape rate) and is connected to the damping time of the reservoir $\tau_B\approx\kappa^{-1}$ which is much longer than its correlation time, over which the correlation functions of the reservoir vanish \cite{Gardiner2004}. On the other hand, the parameter $g$ is related to the spontaneous decay rate of the cavity through which the relaxation time $\tau_R$ over which the state of the system consisting of only one qubit changes is $\tau_R\approx g^{-1}$  \cite{Bellomo2007}.

The Lorentzian distribution (\ref{eq:spcden}) implies the nonperfect reflectivity of the cavity mirrors \cite{breuer2002theory}.  This leads to an exponentially decaying correlation function, with $\kappa$ as the decay rate factor of the cavity as follows:
\begin{equation}\label{eq:solf}
f(t-t_1)=ng^2e^{-\kappa(t-t_1)}e^{-i\Delta(t-t_1)},
\end{equation}
in which, $\Delta=\omega_0-\omega_{\text{qb}}$ is the detuning parameter. We note that, by choosing special values of $\kappa$, it is possible to extract the ideal cavity and the Markovian limits. The former is obtained when $\kappa\rightarrow 0$, which leads to $J(\omega)=ng^2\delta(\omega-\omega_0)$ corresponding to a constant correlation function in which $\delta(\bullet)$ is the usual Dirac delta function. In this situation, the system reduces to a $n$-qubit Jaynes-Cummings model \cite{Tavis1968} with the vacuum Rabi frequency $\Omega_R=\sqrt{n}g$. On the other hand, for small correlation times and by taking $\kappa$ much larger than any other frequency scale, the Markovian regime may be obtained. For the other generic values of $\kappa$, the model interpolates between these two limits.

Next, the Laplace transform technique helps us to solve the integro-differential equation (\ref{eq:diffu}) and obtain the following relation for the surviving amplitude:
 \begin{equation}
{\cal E}(t)=e^{-(i\Delta+\kappa) t/2}\left( \cosh{\left(\Omega t/ 2 \right)} +\dfrac{i\Delta+\kappa}{\Omega}\sinh{\left( \Omega t/ 2\right) }\right),  \label{eq:survivalamplitude}
 \end{equation}
where $\Omega=\sqrt{\kappa^2-\Omega_R^2+2i\Delta\kappa}$, in which $\Omega_R=\sqrt{\Delta^2+4g^2 n}$. The obtained analytical expression for amplitude ${\cal E}(t)$ is exact and therefore outside Markovian regime. Furthermore, at the steady state ($t\longrightarrow \infty$), ${\cal E}(t)\longrightarrow 0$,  looking at (\ref{eq:statewerner})  it turns out that $\ket{\psi(\infty)}\propto \ket{G}$ which means that, the initial entanglement must have a decaying behaviour as time goes on and no stationary entanglement can be achieved. It is worth noticing that for system size $n+1$ the relation derived above reduces the results of the pioneering work on quantum Zeno and anti Zeno effects in a resonator \cite{Kofman1996}. For $W$-states we note the similarity with the result of the general treatment of decoherence control (and the quantum Zeno effect in particular) for entangled states \cite{Gordon2011}. 

It should be noticed that, the exact solution presented in \eqref{eq:survivalamplitude} for strong and weak coupling regime is due to the Lorentzian spectral density which has been directly obtained from our modelling of dissipative cavity. For other kinds of spectral densities, only the weak coupling regime is amenable to a general analysis of the quantum Zeno and anti-Zeno effect \cite{Kofman2000}, where, the authors presented a universal formula for a single qubit weakly coupled to any zero temperature bath under control by non-selective measurements. 
\section{Dynamical Evaluation of Entanglement}\label{sec:OED}
It should be stressed that, due to the approximations of the model, one can consider any pair of quibts.
In the computational basis and using (\ref{eq:statewerner}), the explicit form of the reduced density operator for any pair of qubits, after tracing over the environment degrees of freedom and partial tracing over all other qubits, takes the form
\begin{equation}
\rho_{\text{pair}}(\tau)=\begin{pmatrix}
 0 & 0 & 0 & 0 \\
 0 & \dfrac{\left| {\cal E}(\tau)\right| ^2}{n} & \dfrac{\left| {\cal E}(\tau)\right| ^2}{n} & 0 \\
 0 & \dfrac{\left| {\cal E}(\tau)\right| ^2}{n} & \dfrac{\left| {\cal E}(\tau)\right| ^2}{n} & 0 \\
 0 & 0 & 0 & 1-\dfrac{2\left| {\cal E}(\tau)\right| ^2}{n}
 \end{pmatrix},
 \label{eq:rwkl}
\end{equation}
in which the scaled (dimensionless) time reads as $\tau=\kappa t$. Therefore, the pairwise dynamics is completely characterized by the surviving amplitude of the initial state ${\cal E}(t)$.
In the following, we use concurrence as a suitable measure, ranging from 0 (for separable states) to 1 (for maximally entangled states), to quantify the amount of entanglement between various pairs of qubits, which is defined as \cite{Wootters1998}
\begin{equation}
{\cal C}(t)=\max\left\lbrace 0, \sqrt{\ell_1}- \sqrt{\ell_2}- \sqrt{\ell_3}- \sqrt{\ell_4}\right\rbrace,
\label{eq:con}
\end{equation}
where $\left\lbrace \ell_j\right\rbrace _{j=1}^4$ are the eigenvalues (in decreasing order) of the Hermitian matrix
$\rho\left(\sigma_1^y\otimes\sigma_2^y\rho^{*}\sigma_1^y\otimes\sigma_2^y\right)$ with $\rho^*$ as the complex conjugate of $\rho$ and $\sigma_k^y:=i(\sigma_k-\sigma_k^\dag)$.
 Consequently, the explicit form of the concurrence can be obtained from the reduced density matrix (\ref{eq:rwkl}) as follows
\begin{equation}
{\cal C}_{\text{pair}}(\tau)=\dfrac{2\left| {\cal E}(\tau)\right| ^2}{n}.
\label{eq:conw}
\end{equation}
According to (\ref{eq:suramp}), ${\cal C}_{\text{pair}}(\tau)=2\text{P}_0(\tau)/n$ implies that, the pairwise concurrence directly depends on the survival probability of the initial state.

We recall that, the sequence of $N$  nonselective measurements on the collective of qubits can induce the quantum Zeno effect. In this way, an efficient entanglement protection may be obtained. However, this QZE strongly depends on the environment features and the resonance condition. In some cases, it may cause an enhancement on the decay of entanglement, which is corresponding to the quantum anti-Zeno effect. In order to examine the effect of repeated measurements on the entanglement dynamics, we consider the action of a series of N nonselective measurements, each performed at time intervals $T=t/N$ in order to check whether the system is still in its initial state. 

We assume that the series of nonselective measurements on the collective atomic system, each performing at time intervals $T$, are assumed to have the two following properties: (i) one of its possible outcomes is the projection onto the ground state $\ket{G}:=\ket{0}^{\otimes n}$, and (ii) the measurement cannot distinguish between the states $\ket{1_1,0_2,\cdots,0_k,\cdots,0_n}$, $\cdots$, $\ket{0_1,\cdots,1_k,\cdots,0_n}$, $\cdots$, $\ket{0_1,\cdots,0_k,\cdots,1_n}$. Any procedure fulfilling these two conditions will do the task of measurement. For instance, since we assumed that the transition frequency of all qubits is equal  (consequently, the second condition holds), one can measure the collective atomic energy, which certainly determines that whether the system of qubits have decayed into the ground state $\ket{G}$ or its excitation remains. Another way to do the measurement task is monitoring the state of the cavity. Since only one excitation has been considered, if a photon is added to the cavity, then the qubits have necessarily decayed into the state $\ket{G}$. While, if no photon is found, one can dedicate that the excitation still resides on the qubits.

After every measurement, the system is projected back to its initial state with the probability $\text{P}_0(T)$ and then the temporal evolution starts anew. The survival probability of the initial state after the first observation is $\bra{\psi_0}\rho(T)\ket{\psi_0}=\left| {\cal E}(T)\right| ^2$. The sequence of the N measurements repeatedly brings the system into its initial state with the surviving probability $P_0^{(N)}(t=NT)=\left| {\cal E}(T)\right| ^{2N}$ which can be rewritten after some manipulation as \cite{Facchi2001}
\begin{equation}
P_0^{(N)}(t)=\exp\left[ -\Gamma_z(T)t\right] ,
\label{eq:proN}
\end{equation}
with an effective decay rate $\Gamma_z(T)=-\log\left[ \left| {\cal E}(T)\right| ^2\right]/{T}$. It is obvious that, for a finite time $t=NT$ and in the limit $T\longrightarrow 0$ and $N\longrightarrow\infty$, $\Gamma_z(T)\longrightarrow 0$ and the decay is completely suppressed. It is clear that, the projective measurements not only affect the probability $P_0(t)$, but also modify the time evolution of the entanglement. More explicitly, according to Eqs. (\ref{eq:suramp}), (\ref{eq:conw}) and (\ref{eq:proN}), the modified concurrence becomes
\begin{equation}
{\cal C}_{\text{pair}}^{(N)}(t)=\dfrac{2\exp\left[ -\Gamma_z(T)t\right]}{n}.
\label{eq:modcon}
\end{equation}
This result can also be directly achieved from the density matrix describing the system has been observed $N$ times, i.e., $\ket{\psi(t)^{(N)}}\bra{^{(N)}\psi(t)}$, by tracing over the reservoir degrees of freedom and over all other qubits.
 According to (\ref{eq:modcon}), the effective dynamics of concurrence depends on $T$, system size $n$, the relative coupling between qubits and the cavity field, the cavity damping rate as well as the off-resonance parameter $\Delta$.

\section{Results}\label{sec:RES}
In this section, we intend to examine the role of off-resonance and the sequence of the measurements on the dynamics of the pairwise entanglements. Before that, from  Eq. (\ref{eq:survivalamplitude}), two distinct weak and strong coupling regimes can be distinguished by introducing the dimensionless parameter  $R=g/\kappa$, by which, we are able to analyse our results in two regimes good ($R\gg 1$) and bad ($R\ll 1$) cavities. In the bad cavity limit, the relaxation time is greater than the reservoir correlation time and behaviour of $P_0$ is a Markovian exponential decay. In the absence of detuning and the repeated measurements, the survival probability vanishes faster with system size $n$. In the good cavity limit, the reservoir correlation time is greater than the relaxation time and non-Markovian effects such as revival and oscillation of entanglement become dominant. These effects are due to the long memory of the environment. In the absence of detuning and the repeated measurements, the survival probability has discrete zeros at $t_m=2[ m\pi-\arctan(\Omega_n^{'}/\kappa)]/\Omega_n^{'}$ with $m$ integers and $\Omega_n^{'}=\sqrt{4ng^2-\kappa^2}$.

\subsection{Absence of non-Selective Measurements}\label{sec.asb}
\begin{figure*}[ht]
\centering
\subfigure[\label{Fig2a} \ Bad cavity limit, $R=0.1$.]{\includegraphics[width=0.45\textwidth]{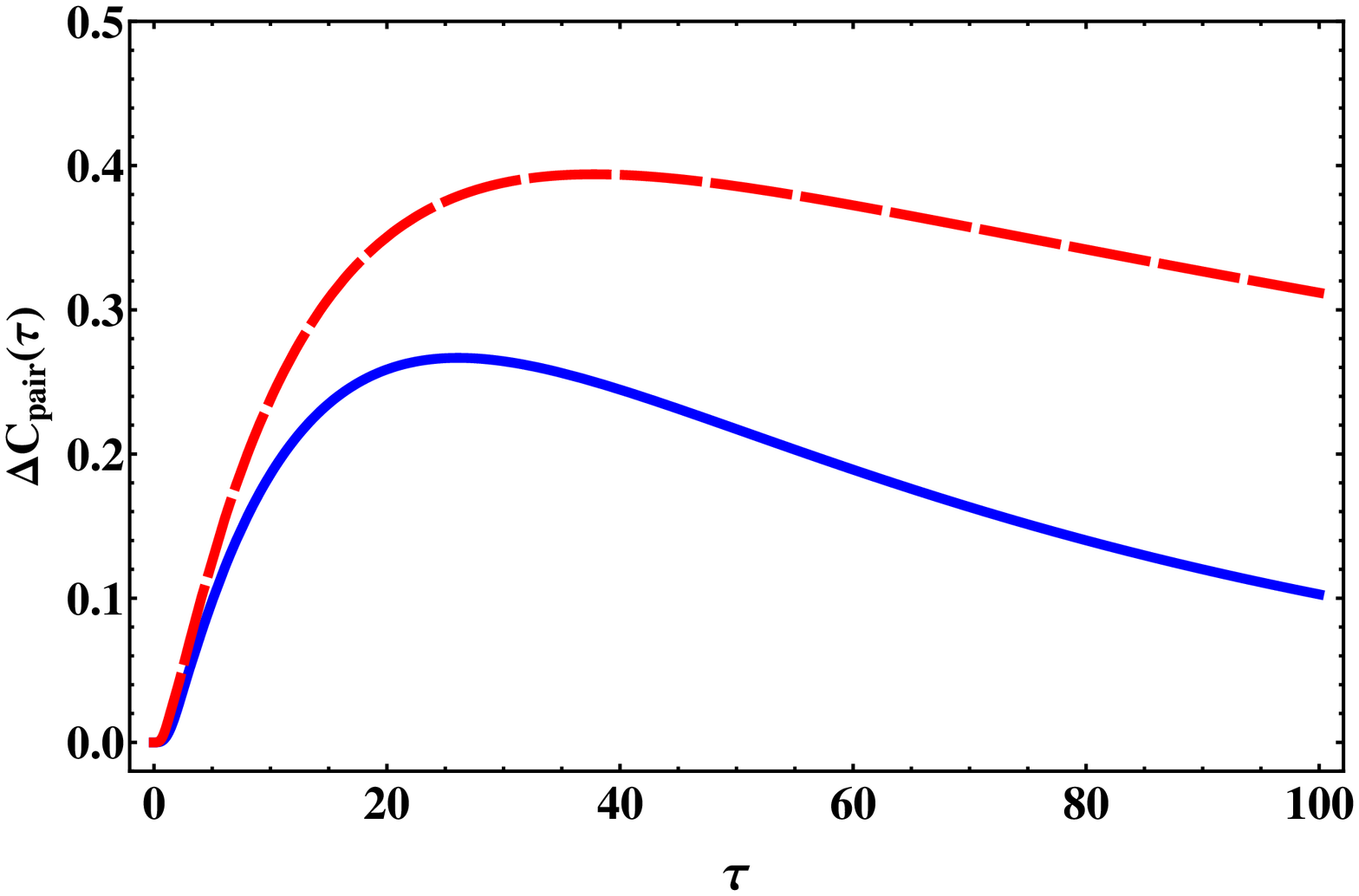}}
\hspace{0.05\textwidth}
\centering
\subfigure[\label{Fig2b} \ Good cavity limit, $R=10$.]{\includegraphics[width=0.45\textwidth]{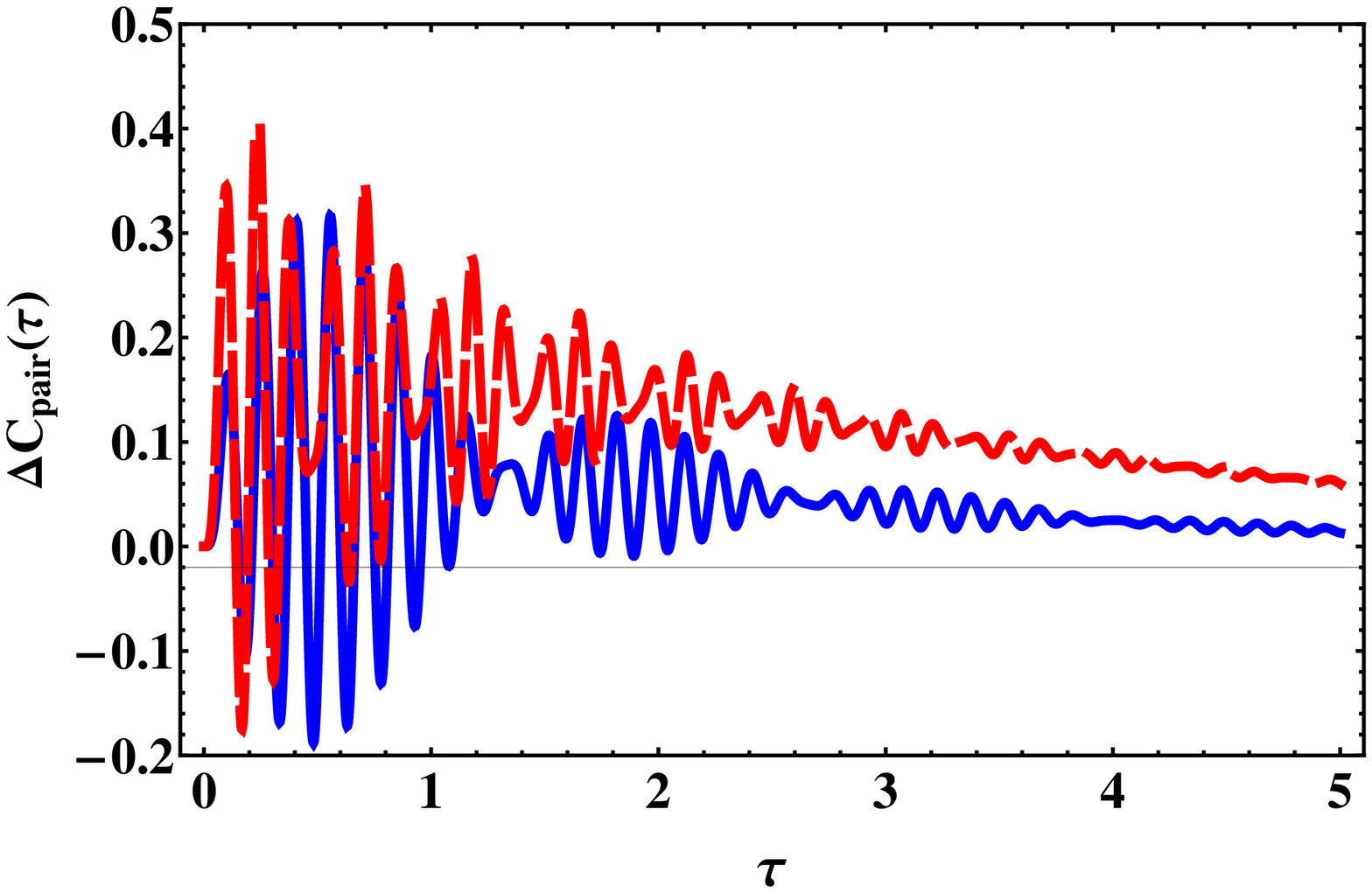}}

\caption{Time evolution of the ${\cal C}_{\text{pair}}(\tau,\Delta)-{\cal C}_{\text{pair}}(\tau,\Delta=0)$ as function of the dimensionless parameter $\tau=\kappa t$ in the absence of repeated measurements for system size $n=4$, (a) in the bad cavity limit, i.e. $R=0.1$ (left plots) for $\Delta=2\kappa$ (blue solid line) and $\Delta=4\kappa$ (red dashed line) and (b) good cavity limit, $R=10$ (right plots) for $\Delta=20\kappa$ (blue solid line)  $\Delta=35\kappa$ (red dashed line).} \label{Fig2}
   \end{figure*}

In Fig. \ref{Fig2}, we have plotted the difference between two concurrences in the presence and in the absence of detuning, i.e., $\Delta{\cal C}_{\text{pair}}(\tau)\equiv{\cal C}_{\text{pair}}(\tau,\Delta)-{\cal C}_{\text{pair}}(\tau,\Delta=0)$ for both strong and weak coupling regimes in the absence of the repeated measurements. As it can be observed, in the weak coupling regime, this quantity is always greater than zero which implies an enhancement in the surviving of the entanglement. Furthermore, according to \ref{Fig2a}, it is possible to achieve maximum enhancement in the surviving of entanglement. This maximum enhancement is sensitive to the detuning parameter $\Delta$ and also the system size $n$. It can be shown that, as the system size increases, this enhancement fades out more quickly. On the other hand, increasing the detuning parameter preserves the pairwise entanglement in longer times under the impact of environmental noise. As is discussed before, the strong regime has a different behaviour. In this regime, an oscillatory behaviour is seen for $\Delta{\cal C}_{\text{pair}}(\tau)$ due to the non-Markovian memory of the reservoir. Therefore, unlike the bad cavity case, in some intervals of time, a decrement in preserving entanglement is seen. However, on average the detuning parameter has a positive role in enhancing the entanglement. Again, increasing the system size deteriorates this possibility. On the other hand, it is possible to achieve a higher enhancement by increasing the detuning parameter. Another difference between these two limits is that, it is possible to achieve a quasi-stationary entanglement in the bad cavity limits by increasing the detuning parameter, but achieving such a state is impossible in the other regime.

\subsection{Presence of non-Selective Measurements with the Exact Resonance Condition}
\begin{figure*}[ht]
\centering
\subfigure[\label{Fig3a} \ Bad cavity limit, $R=0.1$ for $n=2$ with $\Delta=0$.]{\includegraphics[width=0.4\textwidth]{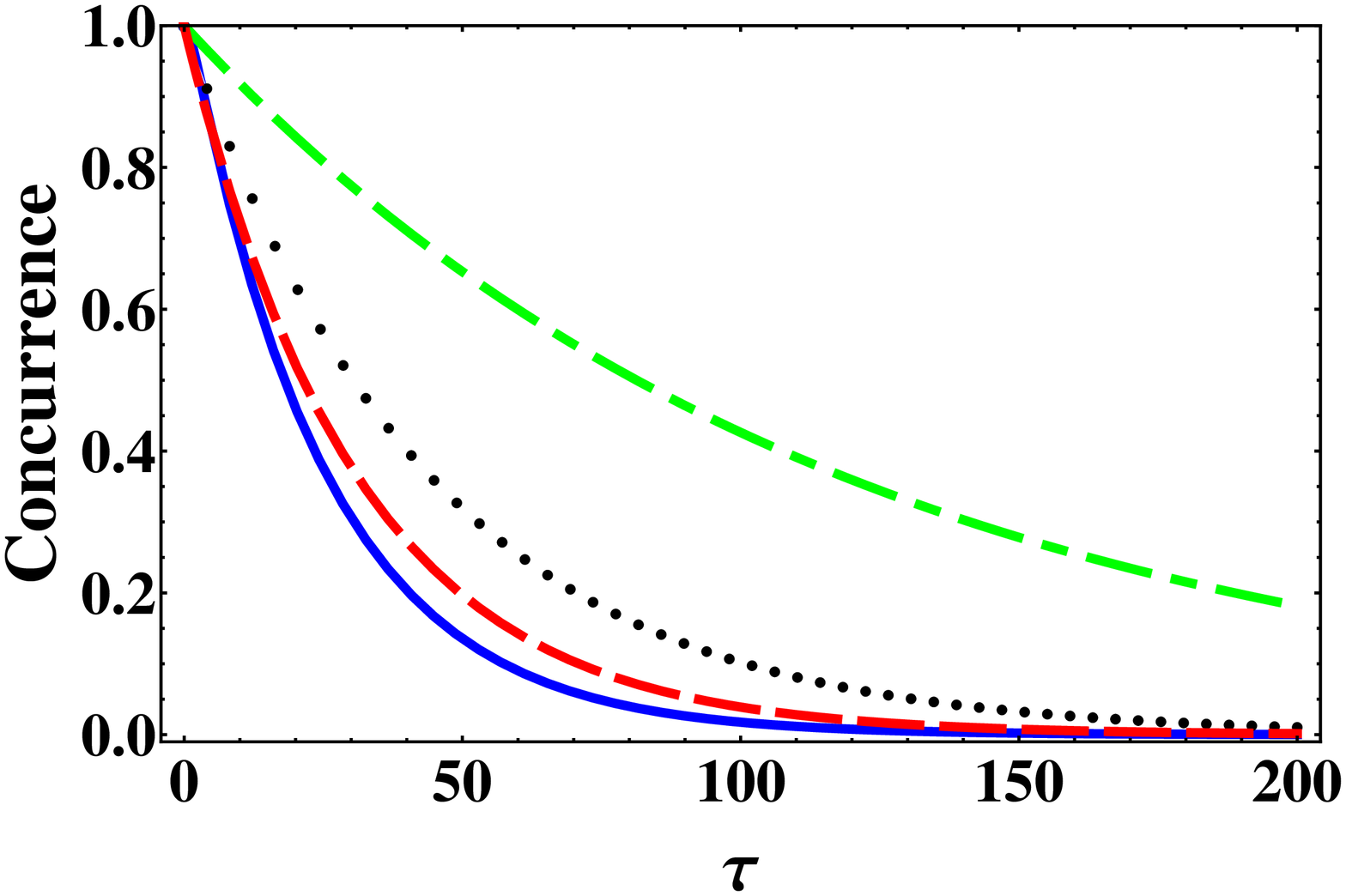}}
\hspace{0.05\textwidth}
\subfigure[\label{Fig3b} \ Good cavity limit, $R=10$ for $n=2$ with $\Delta=0$.]{\includegraphics[width=0.4\textwidth]{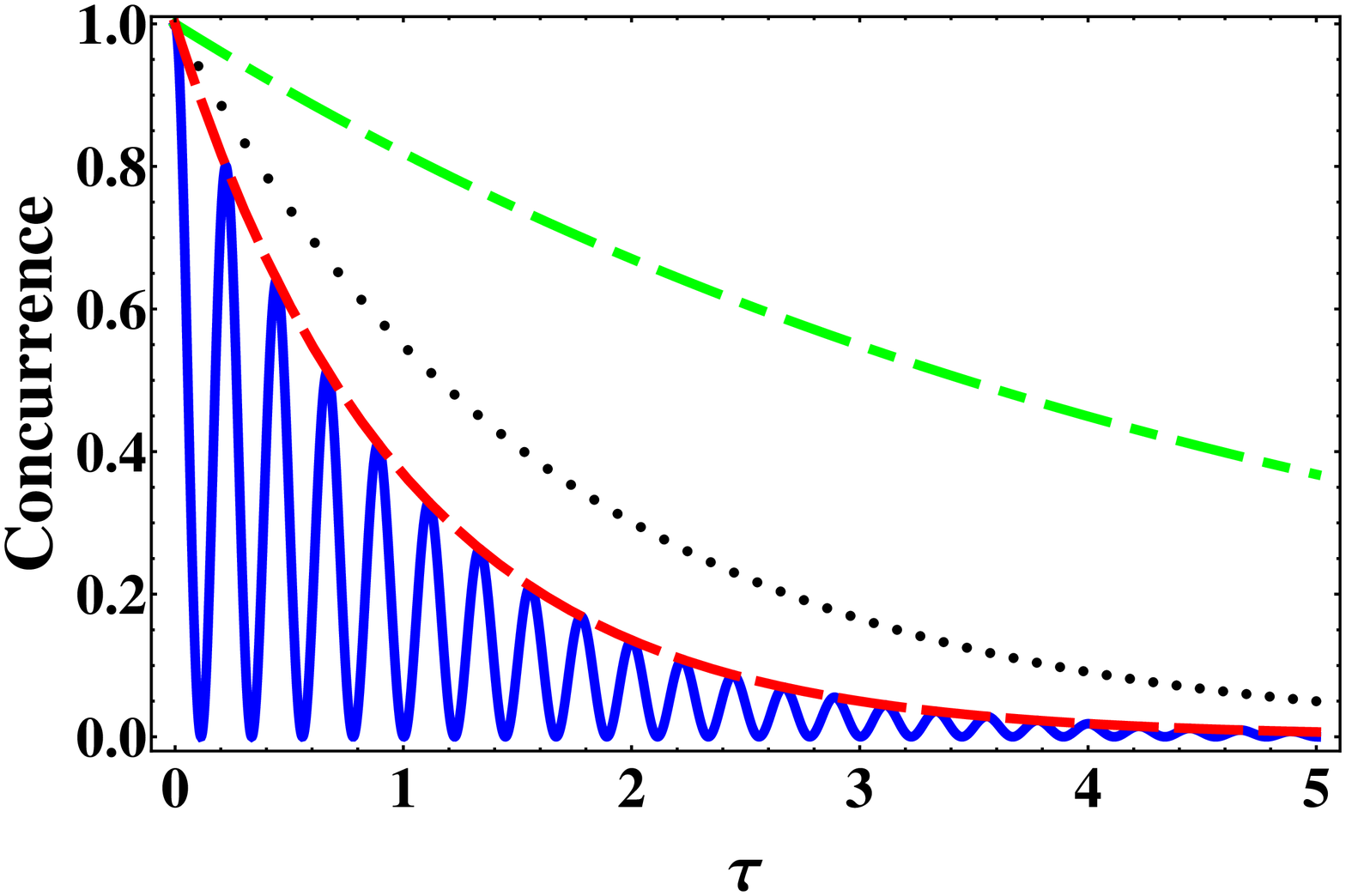}}
\centering
\subfigure[\label{Fig3c} \ Bad cavity limit, $R=0.1$ for $n=4$ with $\Delta=0$.]{\includegraphics[width=0.4\textwidth]{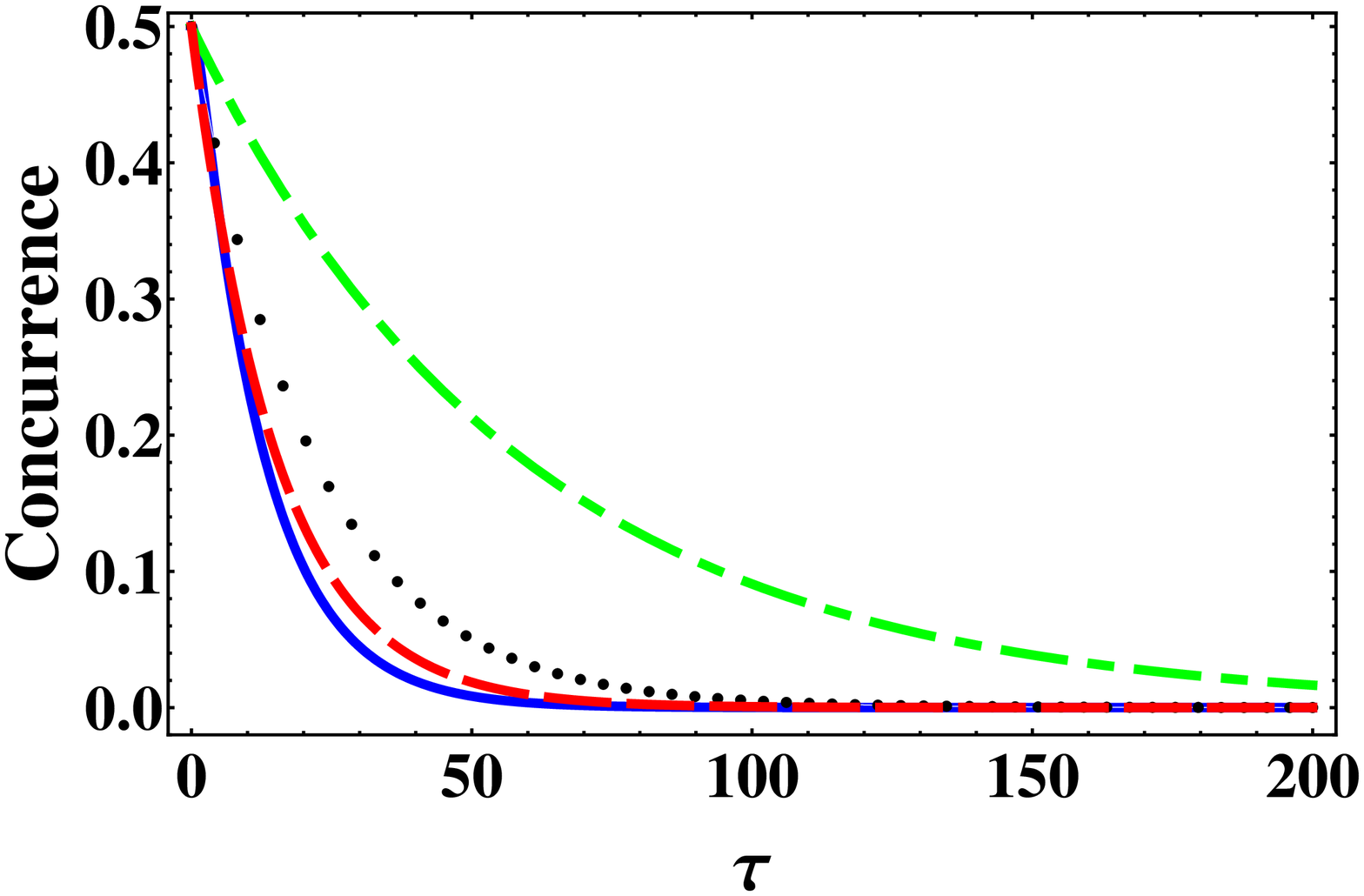}}
\hspace{0.05\textwidth}
\subfigure[\label{Fig3d} \ Good cavity limit, $R=10$ for $n=4$ with $\Delta=0$.]{\includegraphics[width=0.4\textwidth]{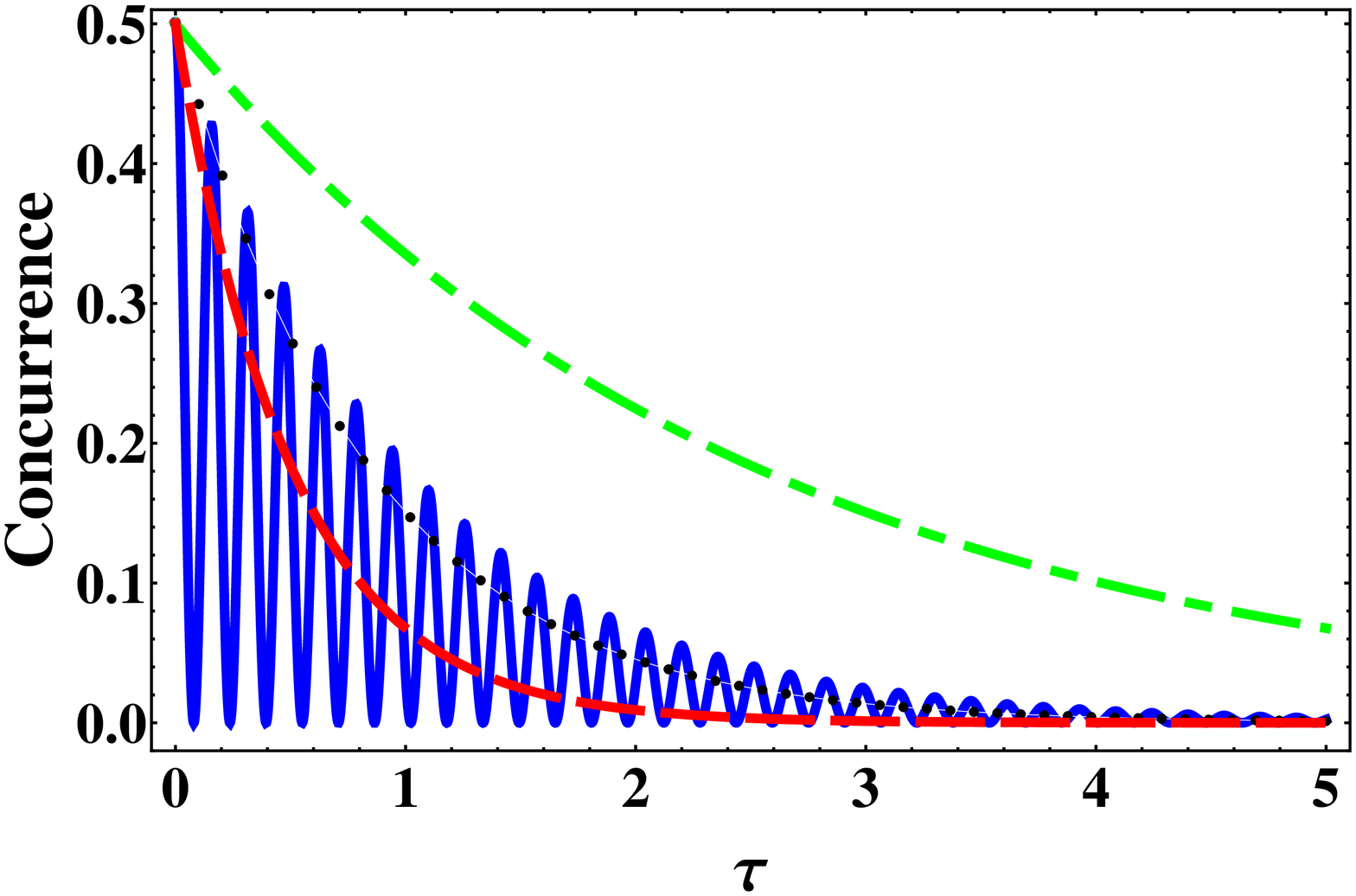}}

\caption{Time evolution of the concurrence as a function of the dimensionless parameter $\tau=\kappa t$ in the absence of detuning $\Delta=0$, for system size $n=2$ (top plots) and $n=4$ (bottom plots), in the absence of measurements (blue solid line) and in the presence of measurements for i) weak coupling ($R=0.1$) (left plots) with intervals $\kappa T=0.5$ (green dotted-dashed line), $2$ (black dotted line) and $5$ (red dashed line) and ii) strong coupling ($R=10$) (right plots) with intervals $\kappa T=0.001$ (green dotted-dashed line), $0.003$ (black dotted line) and $0.005$ (red dashed line).} \label{Fig3}
   \end{figure*}

Figure \ref{Fig3} illustrates the effect of the nonselective measurements on the pairwise entanglement dynamics in the absence of detuning for several values of the system size $n$. In the weak coupling regime, the nonselective measurements quenches the decaying of the pairwise entanglement. This can be shown true for any system size $n$ and any value of the time intervals $T$. Therefore, the observed dynamics shows always the quantum Zeno effect for all values of $n$ and $T$. It is obvious that, by decreasing the system size $n$ and time intervals $T$, entanglement survives in longer times. On the other hand, the strong coupling regime has different behaviour. First of all, in the absence of measurements, the concurrence periodically vanishes according to the zeros of the function $P_0$. These revivals and oscillations are due to the memory depth of the reservoir. Actually, it can be stated that, the reservoir feedbacks part of the information which it has taken during the interaction with the qubits. Performing the repeated measurements on the qubits at time intervals shorter than the reservoir memory time, suppresses the feedback from the reservoir into qubits and then disentangle the qubits from the reservoir and consequently, causes the loss of the oscillatory behaviour of the entanglement.  In the presence of the measurements, the concurrence decreases monotonically to zero and unlike the bad cavity case, the quantum Zeno and anti-Zeno effect may be occurred depending on the system size $n$ and chosen intervals $T$. Moreover, the Zeno region, in which the quenching of the decaying of entanglement occurs, becomes short as the system size $n$ increases.

\subsection{Presence of non-Selective Measurements without the Resonance Condition}
In this section, we intend to examine the role of non-selective measurements and detuning parameter on the entanglement dynamics, simultaneously. This is done in two different coupling regimes. 
\subsubsection{Weak Coupling Regime} 
\begin{figure*}[h!]
\centering
\subfigure[\label{Fig4a} \ Bad cavity limit, $R=0.1$ with $n=4$ and $\Delta=0.5\kappa$.]{\includegraphics[width=0.45\textwidth]{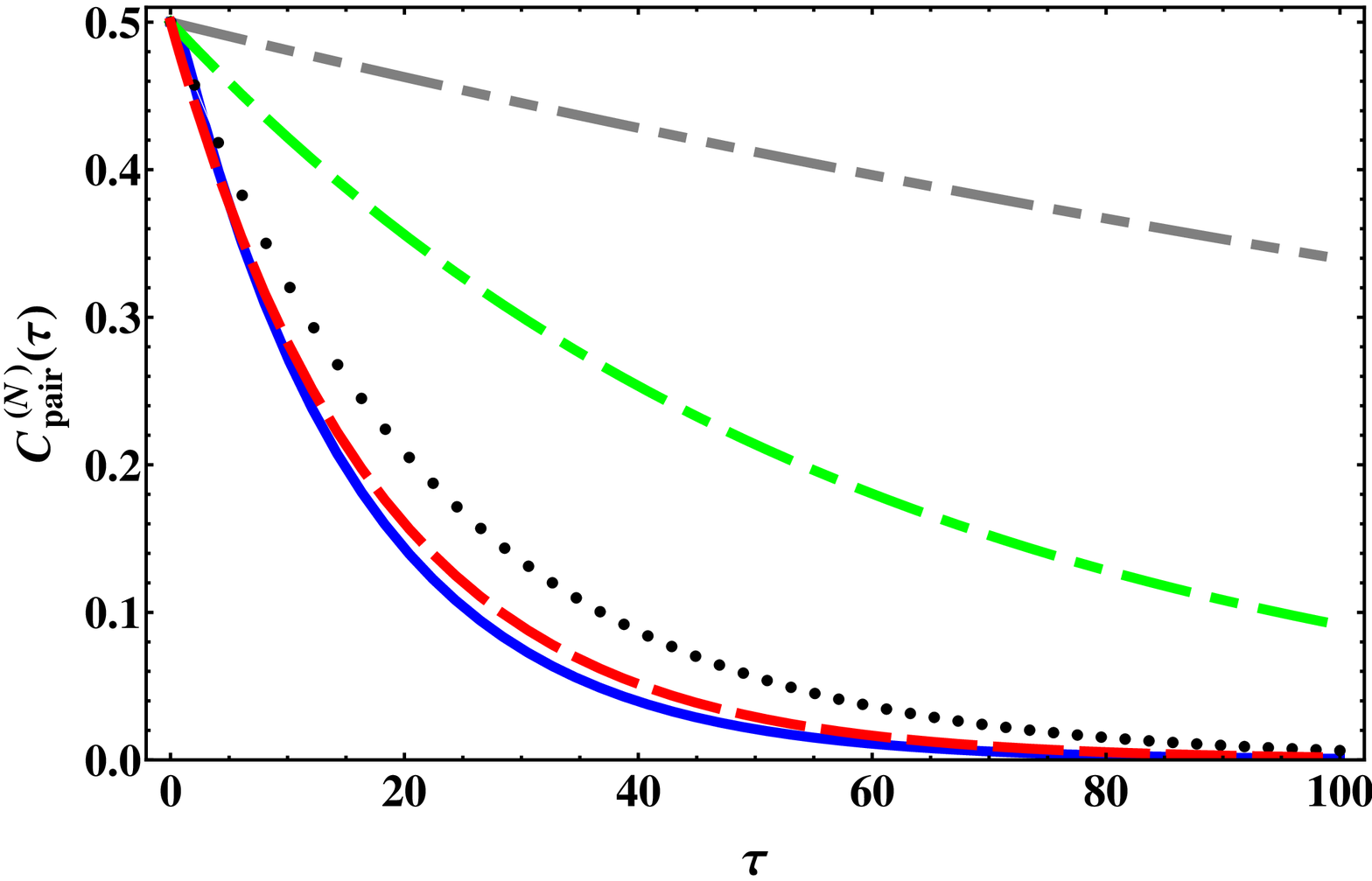}}
\hspace{0.05\textwidth}
\centering
\subfigure[\label{Fig4b} \ Bad cavity limit, $R=0.1$ with $n=4$ and $\Delta=2\kappa$.]{\includegraphics[width=0.45\textwidth]{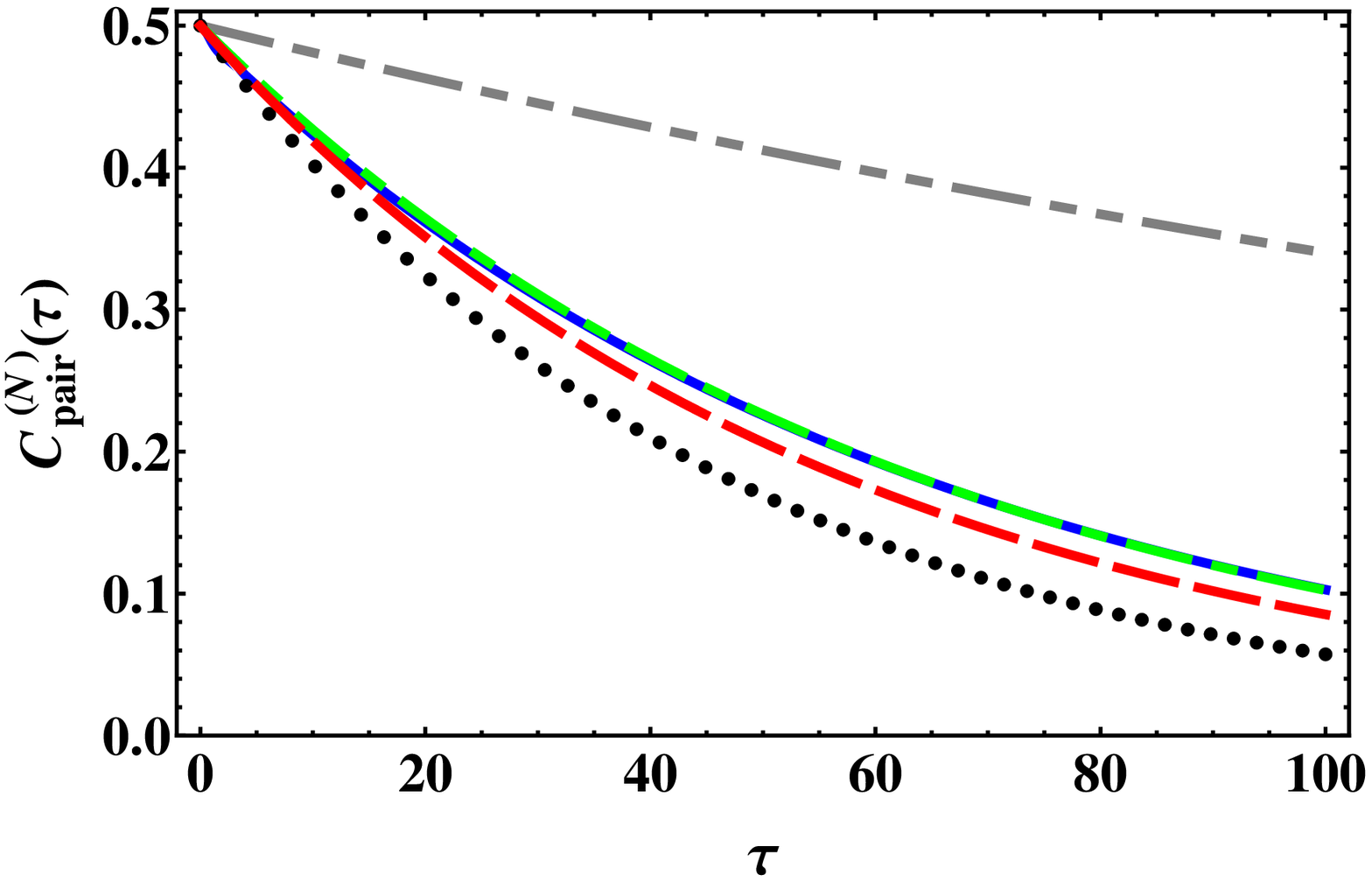}}
\caption{Time evolution of  ${\cal C}_{\text{pair}}^{(N)}(\tau,\Delta)$ as function of the dimensionless parameter $\tau=\kappa t$ in the bad cavity limit, i.e. $R=0.1$ for system size $n=4$, in the absence of repeated measurements (blue solid lines) and in the presence of measurements performed at time intervals: $\kappa T=0.1$ (gray dot-dot-dashed lines), $\kappa T=0.5$ (green dot-dashed lines), $\kappa T=2$ (black dotted lines) and $\kappa T=5$ (red dashed lines) for (a) $\Delta=0.5\kappa$ (left plots)  and (b) $\Delta=2\kappa$ (right plots).} \label{Fig4}
   \end{figure*}

Figure \ref{Fig4} shows the pairwise concurrence in the presence of the detuning parameter and the nonselective measurements in the bad cavity limit, i.e., $R=0.1$ for system size $n=4$. In the dispersive regime, i.e. for values of the detuning $\Delta\leq\kappa$ and in the presence of the nonselective measurements, the detuning does not affect the behaviour of concurrence appreciably compared to the resonant case (compare Figs. \ref{Fig3c} and \ref{Fig4a}). Therefore, in this case the Zeno effects are dominant for any value of the time intervals $T$. This property can also be proven to be true for any value of the system size $n$. On the other hand, by considering the detuning parameter greater than the cavity damping rate $\kappa$ (i.e., $\Delta>\kappa$), the quantum anti-Zeno effect appears for values of $T$ greater than a characteristic threshold value $T^*$ \cite{Facchi2001}. This threshold time depends only on the detuning parameter and does not depend on the system size $n$, such that, for greater values of detuning parameter, this threshold time decreases. The interesting aspect here is that, for the time intervals greater but near to this threshold time, the concurrence vanishes faster. But, increasing the time intervals $T$, the quantum anti-Zeno effect becomes less and less dominant. It can also be shown that, for greater values of the detuning parameter, the Zeno region becomes smaller and it occurs only for very short measurements time intervals.

\subsubsection{Strong Coupling Regime} 
\begin{figure*}[h!]
\centering
\subfigure[\label{Fig5a} \ Good cavity limit, $R=10$ with $n=4$ and $\Delta=5\kappa$.]{\includegraphics[width=0.45\textwidth]{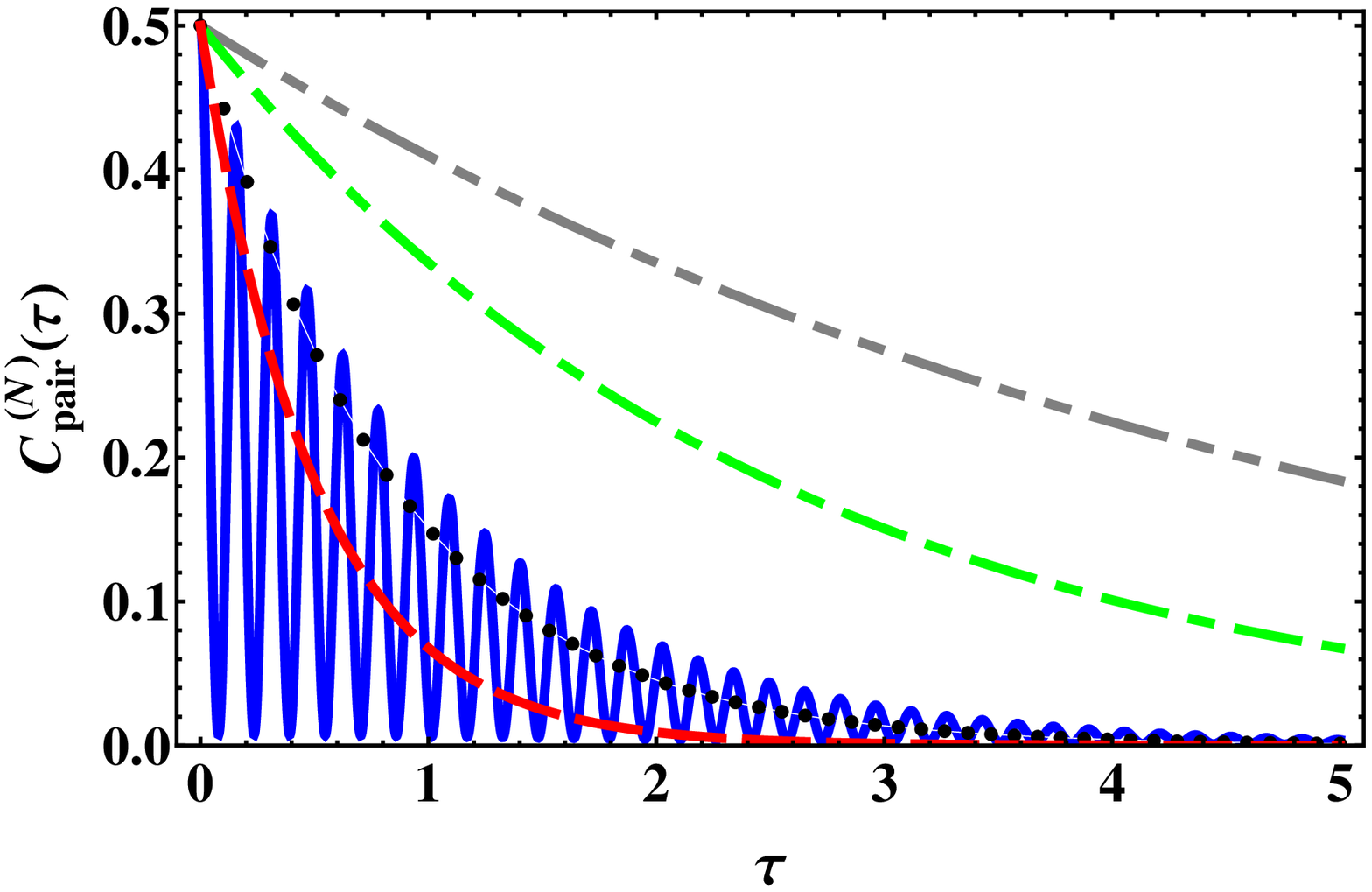}}
\hspace{0.05\textwidth}
\centering
\subfigure[\label{Fig5b} \ Good cavity limit, $R=10$ with $n=4$ and $\Delta=20\kappa$.]{\includegraphics[width=0.45\textwidth]{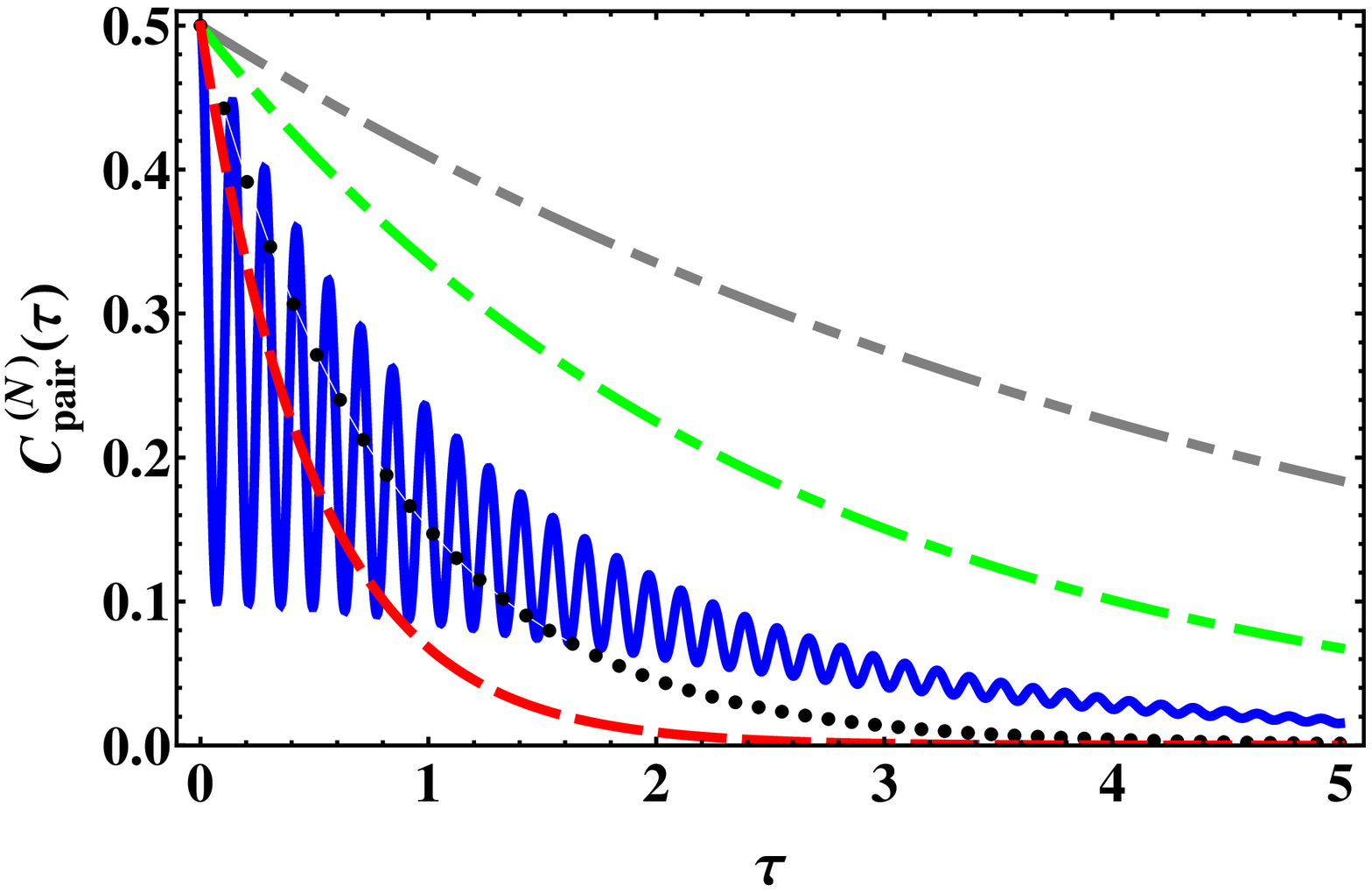}}
\caption{Time evolution of  ${\cal C}_{\text{pair}}^{(N)}(\tau,\Delta)$ as function of the dimensionless parameter $\tau=\kappa t$ in the good cavity limit, i.e. $R=10$ for system size $n=4$, in the absence of repeated measurements (blue solid lines) and in the presence of measurements performed at time intervals: $\kappa T=0.0005$ (gray dot-dot-dashed lines), $\kappa T=0.001$ (green dot-dashed lines), $\kappa T=0.003$ (black dotted lines) and $\kappa T=0.005$ (red dashed lines) for (a) $\Delta=5\kappa$ (left plots)  and (b) $\Delta=20\kappa$ (right plots).} \label{Fig5}
   \end{figure*}

In Fig. \ref{Fig5} we have plotted the pairwise concurrence as a function of $\tau$ for system size $n=4$ in the presence of the detuning and also repeated measurements in the good cavity limit ($R=10$). For small values of detuning, i.e. $\Delta<g$, the concurrence has nearly the same behaviour as resonance case for all values of system size $n$. Therefore, in this case the quantum Zeno and anti-Zeno effects can dominate the dynamics of entanglement depending on the system size $n$ as well as time intervals $T$. The same holds true for $\Delta\sim g$. In the other regime, i.e., $\Delta>g$, the detuning affects considerably the concurrence in the absence of the nonselective measurements. The entanglement sudden death is completely disappeared. In this case the anti-Zeno effect may occurs for values of $T$ greater than a characteristic threshold value $T^*$ \cite{Facchi2001}. Unlike the bad cavity limit, the threshold time not only depends on the detuning parameter but also on the system size $n$. The Zeno region decreases with increasing the system size $n$ and also the detuning parameter. In the absence of the measurements,  for larger values of detuning, the amplitude of the oscillations decreases until the concurrence reaches a monotonically decaying behaviour.

\section{Summary and Conclusion}\label{sec:CON}

To sum up, we have evaluated the dynamics of pairwise entanglement of a collective of an arbitrary number of non-interacting qubits initially prepared in a Werner state, dissipating into a common and non-Markovian environment in the absence and presence of a series of nonselective measurements performing at time intervals $T$ to check whether the system is in its initial state or not. We then investigated the role of detuning parameter in the dynamics of entanglement. We also found the conditions for both the quantum Zeno and anti-Zeno effects on the pairwise entanglement. We found that in the absence of measurements and detuning, when qubits are initially in a Werner state, the concurrence has a monotonically decaying (in the bad cavity limit) or an oscillatory decaying behaviour (in the good cavity limit) without any stationary value. This is in consistent with previous works when two qubits are initially prepared in a Bell state and dissipate into a common environment \cite{Nourmandipour2015,Maniscalco2008}. In the absence of repeated measurements, the detuning parameter has always a positive role in surviving of initial entanglement in the bad cavity limit. It is even possible to achieve a stationary entanglement in the weak coupling regime by increasing the values of detuning parameter. But, in the good cavity limit, due to the fact that the long memory of the environment can induce the oscillations and revivals of entanglement, we have a decrement in preserving of entanglement in some chosen values of time intervals. In the absence of detuning and in the bad cavity limit, the observed dynamics shows always the quantum Zeno effect for any value of the system size $n$ and time intervals $T$. For smaller values of $n$ and  $T$, the quantum Zeno effect is stronger. In the good cavity limit, the quantum Zeno effect can not be easily predictable, since it depends on the system size, the detuning parameter and also on the measurement times $T$. Therefore, the quantum anti-Zeno effect may also occur for values of $T$ greater than a threshold time $T^*$ which depends on the system size $n$. The behaviour of concurrence in the bad cavity limit and in the presence of nonselective measurements for detuning parameters less than the cavity damping rate (i.e., $\Delta\lesssim\kappa$) is similar to the resonance case and the quantum Zeno effect is always dominant. But, for values of detuning parameter greater than the cavity damping rate, the anti-Zeno effect may also occur. On the other hand, in good cavity limit and for any values of detuning parameter, the quantum Zeno and anti-Zeno effects can appear depending on the measurements times and the system size.

It is worth noticing that our results are completely in consistent with previous works. For instance, for  system size $n=2$, our results reduce to the ones presented in \cite{Francica2009} for the subradiant scenario, i.e, $\omega_{\text{qb}_1}=\omega_{\text{qb}_2}$. This motivates us to investigate the quantum Zeno and anti-Zeno effects on the quantum and classical correlations for an arbitrary number of qubits \cite{Francica2010}, too, which is left for our future works.

Finally, we should notice that, our results  could be verified and confirmed in experiments relating to the trapped ions coupled to the dissipative bath of vacuum modes of the radiation field via optical pumping \cite{Barreiro2011}. In addition, the system of superconducting Josephson circuits as qubits and a transmission line as cavity could be a suitable candidate for an experimental implementation which may explore the contents of the paper \cite{Majer2007}. The present work can also be relevant for driving cavity QED experiments with noninteracting qubits inside a cavity \cite{Specht2011}.


\end{document}